\documentclass[letterpaper]{article}
\pdfoutput=1
\usepackage{aaai}
\usepackage{graphicx}
\usepackage{epstopdf}
\usepackage{amsmath}
\usepackage{amsfonts}
\usepackage{algorithmic,algorithm}

\begin{document}

\title{Knowledge-based Query Expansion in \\ Real-Time Microblog Search}
\author{
Runwei Qiang \quad
Feifan Fan \quad
Chao Lv \quad
Jianwu Yang\thanks {Corresponding author.}\\
\{qiangrw, fanff, lvchao, yangjw\}@pku.edu.cn\\
Institute of Computer Science and Technology\\
Peking University,Beijing 100871,P.R.China
}
\nocopyright
\maketitle
\begin{abstract}
Since the length of microblog texts, such as tweets, is strictly limited to 140 characters, 
traditional Information Retrieval techniques suffer from the vocabulary mismatch problem severely
and cannot yield good performance in the context of microblogosphere.
To address this critical challenge, in this paper, 
we propose a new language modeling approach for microblog retrieval 
by inferring various types of context information.
In particular, we expand the query using knowledge terms derived from Freebase 
so that the expanded one can better reflect users' search intent.
Besides, in order to further satisfy users' real-time information need, 
we incorporate temporal evidences into the expansion method,
which can boost recent tweets in the retrieval results with respect to a given topic.
Experimental results on two official TREC Twitter corpora 
demonstrate the significant superiority of our approach over baseline methods.
\end{abstract}
\section{Introduction}
Information Retrieval (IR) in the microblogosphere such as Twitter \footnote{https://www.twitter.com}
has attracted increasing research attention along with the fast development of social media.
To explore the information seeking behavior in microblogoshpere,
TREC first introduced a Real-Time Search Task (RTST) in 2011 \cite{ounis:trec11},
which can be summarized as ``At time $T$, give me the most relevant tweets about topic $X$".

However, it is inherently challenging to develop an effective real-time IR platform 
in the context of microblogosphere.
First, in contrast to traditional web search techniques, 
real-time search task usually faces the problem of severe vocabulary mismatch.
Since the tweets are very short, 
there is a large risk that query terms fail to match any word observed in relevant tweets. 
This problem is extremely severe especially when people search the entities with several alternative aliases.
Besides, real-time search usually indicates the information need of something happening right now.
Thus, it is very crucial for the IR approach to favor the recent tweets relevant to the given topic. 
This real-time information need requires search engines to trade off 
between the recency and relevance score computed between the query and tweet.

Query Expansion (QE) methods based on pseudo-relevance feedback (PRF) 
\cite{liang:jcdl12,lv2009comparative,zhai2011mbfb} 
are widely used in microblog search to mitigate the problems mentioned above.
However, these methods rely much on the assumption that the top ranked documents 
in the initial search are relevant and contain good words for query expansion. 
Nevertheless, in real world, 
this assumption does not always hold in microblogosphere \cite{cao2008selecting,miyanishi2013improving}, 
considering the example that the query contains proper nouns difficult to understand.
What's more, even if the top ranked documents are highly relevant to the topic, 
it is still very likely that they contain numerous topic-unrelated words 
due to the informality of the tweet content \cite{miyanishi2013improving}.

To overcome the limitations of existing methods, 
we utilize Freebase\footnote{http://www.freebase.com} as the knowledge source
to infer more topic-related context information for each query.
Freebase is a practical, scalable tuple database used to organize general human knowledge \cite{bollacker2008freebase},
covering a large amount of knowledge in different aspects (domains), into a hierarchical structure.
In contrast to Wikipedia that describes human knowledge with long detailed articles 
and WordNet that mainly contains synonymy relations, 
Freebase represents the human knowledge using an ontological structure (i.e. \textit{types}).
Different types, including alias, notable\_for and description, provide different data views for each specific concept.
In this paper, we propose a knowledge query generation method,
in which we first match related concepts in Freebase with respect to the query,
and then extract useful terms from different properties of the concepts to generate the knowledge query.
By interpolating the original query with the knowledge query, 
we can better reflect the users' information need.

To further utilize the temporal evidence in microblogosphere,
we follow the work of \cite{li2003time} and incorporate a prior distribution 
regarding to the recency of documents into the language modeling frameworks.
More specifically,
while selecting top knowledge terms from Freebase using an association based method,
we assign each top ranked pseudo-relevance document with a time prior so that the words 
appearing more in recent documents are associated with higher probability.

The main contributions of this paper include:
(1) we propose a novel approach to generate knowledge terms from Freebase to expand the original query, 
which leads to better understanding of information need;
(2) the temporal evidence is incorporated into our QE method to trade off between relevance and recency;
(3) we perform a set of experiments on two official twitter test collections published by TREC,
to compare our proposed method with the state-of-the-art baseline methods.
And, the experimental results demonstrate that our proposed approach can give rise to significant better retrieval performance.

\section{Related Work}
\subsection{PRF-based Query Expansion}
QE methods based on PRF assume that most frequent terms in the pseudo-relevance documents are useful, 
which may not always hold in practice.
\cite{cao2008selecting} then integrated a term classification process 
to predict the effectiveness of expansion terms.
\cite{miyanishi2013improving} proposed a manual tweet selection feedback (TSF)
to improve the retrieval performance. 
They further used a two-stage PRF based on similarity of temporal profiles of the query and top retrieved tweets.
However, this method sometimes fails due to the content redundancy of tweets, 
which contain meaningless words that may degrade search results.
Thus, to improve the retrieval performance more using TSF, 
they suggest to detect important concepts from the feedback tweet.

\subsection{Knowledge-based Query Expansion}
Several approaches have been proposed to use the external resource such as Wikipedia, WordNet and ConceptNet to
improve query expansion \cite{collins2005query,xu2009query,kotov2012tapping}.
\cite{li2007improving} explored the possibilities of using Wikipedia's articles
as an external corpus to expand ad-hoc queries
and demonstrated that Wikipedia especially useful to improve weak queries which PRF is unable to improve.
In their methods, expansion terms were extracted from the top ranked Wikipedia articles.
To fill the gap between vocabularies used in indexed documents and user queries,
\cite{aggarwal2012query} used Wikipedia to retrieve the K-best related concepts to the query.
They utilized Wikipedia and and DBPedia to generate the concept candidates, 
and then ranked  them according to the semantic relatedness score given by the Wikipedia-based Explicit Semantic Analysis (ESA) \cite{gabrilovich2007computing}.
\cite{pan2013using} proposed using Dempster-Shafer's Evidence Theory 
to measure the certainty of expansion terms from the Freebase structure.
To the best of our knowledge, query expansion based on Freebase knowledge 
in microblog search is novel and effective.
Unlike previous works, 
our method explored the related concepts in Freebase and attempted to find their aliases 
to solve the vocabulary mismatch problem.
Besides, an association based term selection method is adopted to select useful expansion terms to better understand the users' search intent.

\subsection{Temporal Evidence}
Previous works showed that temporal evidence can be incorporated into IR
\cite{dakka2012answering,dong2010time}.
\cite{li2003time} exploited a prior distribution 
regarding to the recency of documents in the language modeling frameworks for retrieval.
\cite{liang:jcdl12} proposed a temporal re-ranking component
to evaluate the temporal aspects of documents.
\cite{efron2011estimation} proposed IR methods using temporal property in language modeling and showed their effectiveness for recency queries.
\cite{miyanishi2013improving} assumed that similar temporal models share similar temporal property
and proposed a query-document dependent temporal relevance model.
\cite{albakour2013sparsity} introduced a decay factor 
to balance the short-term and long-term interests for a given topic.
In our study,  the temporal evidence is well incorporated in the expansion method 
in order to enhance the importance of the words those are often used to describe the concept recently.

\section{Proposed Methods}
Given the RTST,
we assume that a query $Q$ is obtained as a sample from a generative model $\hat\theta_Q$,
while the document $D$ is generated by model $\hat\theta_D$.
If $\hat\theta_Q$ and $\hat\theta_D$ are the estimated query and document language model
respectively, according to \cite{lafferty2001document},
the relevance score of $D$ with respect to $Q$ can be computed by the following negative KL-divergence function:
\begin{equation}
    S(Q,D) = -D(\hat\theta_Q||\hat\theta_D) \propto \sum_{w \in V} P(w|\hat\theta_Q)\cdot\log{P(w|\hat\theta_D)}
    \label{kl}
\end{equation}
Within this ranking formula, the retrieval problem is essentially equivalent to the problem
of estimating $\hat\theta_Q$ and $\hat\theta_D$.
In principle, we can use any language model for the query and document, which is very flexible.

The start point of our study is to infer more topic-related context for the query with the help of Freebase.
In this section, we first elaborate on why we choose Freebase as our knowledge base.
Based on the characteristics of Freebase,
we describe our proposed method of knowledge query generation in detail.
Further improvements can be obtained by combining the knowledge-based query expansion 
with model-based pseudo-relevance feedback method.

\subsection{Why We Choose Freebase}
Freebase is a large collaborative knowledge base consisted of data harvested from
sources such as the Semantic Web and Wikipedia,
as well as individually contributed data from community members \cite{bollacker2007freebase}.
In Freebase, human knowledge is described by structured categories,
which are also known as \textit{types} and each type has a number of defined \textit{properties}.
In this way, Freebase merges the scalability of structured databases with the diversity of collaborative wikis
into a structured general human knowledge \cite{bollacker2008freebase}.
Just as properties are grouped into types, types themselves are grouped into domains. 
Domains can be considered as the sections in your favorite newspaper: Business, Life Style, Arts and Entertainment, Politics, Economics, etc.

Table \ref{tab:fbeg} shows some topic types of Freebase concept ``Mila Kunis'' in common domain.
As we can see from it, most types provide us useful information to understand the concept ``Mila Kunis'' and
thus can be used as knowledge context of the original concept.
This structured knowledge shows two superiorities compared with the semi-structured or plain contents:
\begin{enumerate}
\item When searching in the Freebase (with API), different types can be integrated for a more accurate concept,
and some types such as name and alias are more important;
\item When generating knowledge terms, we can treat different types and the corresponding properties as different
evidence sources.
\end{enumerate}

\begin{table}[htbp!]
    \centering
    \caption{Common topic types of Freebase concept ``Mila Kunis''.}
    \begin{tabular}{|p{56pt}|p{164pt}|}
        \hline
        \textbf{Type} & \textbf{Property} \\
        \hline
        name & Mila Kunis \\
        \hline
        alias & Milena Markovna Kunis, $\ldots$ \\
        \hline
        notable\_for & Actor \\
        \hline
        notable\_types & Celebrity \\
        \hline
	 description & Milena Markovna is an American actress and voice artist. In
        1991, at the age of seven, she moved from the Soviet Union to Los Angeles with her family $\ldots$ [Summary Description From Wikipedia] \\
        \hline
    \end{tabular}
    \label{tab:fbeg}
\end{table}
\begin{table*}[htbp]
\centering
    \caption{Top retrieved tweets for TREC topics.}
    \begin{tabular}{|p{50pt}|p{120pt}|p{300pt}|}
        \hline
        \textbf{Topic No.} & \textbf{Topic} & \textbf{Relevant Tweet Example} \\
        \hline
        MB071 & Australian Open Djokovic vs. Murray & Tomorrow is the Australian open tennis final for men, Andy Murray vs. Navok Djokovic Who’s gonna win?? I’m a Murray fan so I say GO MURRAY!!” \\
        \hline
        MB115 & memories of Mr. Rogers  & ``@MellowAnniston: Happy late Birthday to Mr. Rogers! '' Omg, Mr. Rogers and I have the same bday? Lol \\ 
        \hline
        MB141 & Mila Kunis in Oz movie & Aw new Oz movie why you go make Mila Kunis ugly, why sir WHY?! \\
        \hline
        MB150 & UK wine industry &  Wine, grape industry accounts for \$6.8bn in Canadian economy: Report: Wine and grape industry in Canada accoun ...  \\
        \hline
    \end{tabular}
    \label{tab:prt}
\end{table*}

Moreover, unlike Twitter including many meaningless and topic-unrelated terms 
(See terms used in some top retrieved tweets for TREC topics in Table \ref{tab:prt}),
the terms used in Freebase is always quite formal and semantically related with the specific concept.
Thus, we assume that the utilization of knowledge terms for query expansion can be more effective 
to improve the overall retrieval performance.

\subsection{Generation of Knowledge Query}
We generate the knowledge query based on types,
aiming at extracting terms from different properties for a given query.
The basic procedures of our proposed method include:
\begin{itemize}
\item \textbf{Concept Match.}
    We select the topic-related concepts with the help of Freebase API.
    Taking the query ``Mila Kunis in Oz Movie'' (MB141) as an example, we match two concepts ``Mila Kunis'' and
    ``The Wizard of Oz'' in Freebase.
\item \textbf{Term Selection.}
	Freebase describes the human knowledge of a given concept using types and properties. 
	For some important meta types such as alias, name, notable\_for and notable\_types in common domain, 
	we directly add terms from these corresponding properties to the knowledge query.
	For other types (i.e. description and domain specific types), 
	we adopt an association based term selection method to extract the topic-related top $K$ terms.
	
	Taking the concept ``Mila Kunis'' as an example,
      we can conduct the term selection methods to gain the top knowledge words ``oz, great, power'' from description property,
      and directly add knowledge terms ``celebrity actor milena markovna kunis'' from meta properties.
\end{itemize}

Then, we view the selected knowledge terms from all related concepts equally
to form a new knowledge query $Q_{fb}$.
After that, the knowledge query model $\hat\theta_{Q_{fb}}$ is interpolated with the original query model $\hat\theta_Q$:
\begin{equation}
    P(w|\hat\theta_{Q_1}) = (1-\alpha) \cdot P(w|\hat\theta_Q) + \alpha \cdot P(w|\hat\theta_{Q_{fb}})
    \label{interpolate}
\end{equation}
where $\alpha \in [0,1]$ is the weighting parameter to control the influence of the knowledge query.
Both $\hat\theta_Q$ and $\hat\theta_{Q_{fb}}$ are estimated according to the maximum likelihood estimator.

\subsubsection{Concept Match}
\label{sec:cm}
We then describe our concept match algorithm in detail, which can be concluded as two steps:
\begin{enumerate}
    \item \textbf{Noun Phrase Detection.}
        For a given query $Q$, we first split $Q$ by space and receive a sequence of words $q_1, q_2, \cdots q_n$.
        Part-of-speech Tagging \cite{RothZe98} is then performed on each word,
        and all the noun phrases are extracted with rule-based method \cite{bird2009natural} from the original query.
    \item \textbf{Maximum Match.} For each noun phrase,
        we regard it as a new query and get the related concepts as described in \textbf{Algorithm} \ref{alg:mm}.
        The FreebaseSearch function searches the given query in the Freebase and returns the
        top ranked concept if found.
        The match process ends if a related concept is found or none of the separate words can find a match.
	  For the sake of efficiency,
	  we can use a hash map which records the searched substrings to avoid duplicating call of FreebaseSearch.
\end{enumerate}
\begin{algorithm}[htpb]
\caption{GetConcept(NQ)}
\label{alg:mm}
\begin{algorithmic}[1]
\REQUIRE ~~\\
Noun Phrase Query $NQ = q_1q_2 \cdots q_n$.
\ENSURE ~~\\
Candidate Concept Set $CSet$.
\STATE $CSet$ = FreebaseSearch($NQ$)
\IF {$CSet$ is empty}
\IF {n == 1}
\RETURN $\emptyset$
\ENDIF
\STATE $NQ_1 \leftarrow q_1q_2 ... q_{n-1}$
\STATE $NQ_2 \leftarrow q_2q_3 ... q_{n}$
\STATE $CSet_1$ $\leftarrow$ GetConcept($NQ_1$)
\STATE $CSet_2$ $\leftarrow$ GetConcept($NQ_2$)
\RETURN $CSet_1 \cup CSet_2$
\ELSE
\RETURN $CSet$
\ENDIF
\end{algorithmic}
\end{algorithm}

\subsubsection{Term Selection}
For each returned concept from Freebase API,
different types and corresponding properties 
which reflect the different aspects of the concept are provided by the search result.
Some types (i.e. meta types) are very general and precise, such as alias, name, notable\_for and notable\_types in the common domain,
we directly add the property terms to the knowledge query for these types.
When it comes to other types such as description and domain specific ones with long texts,
an association based term selection method is utilized to extract the topic-related knowledge terms.
Effective term selection is an important issue for an automatic query expansion technique.
In microblog retrieval, a good expansion term should satisfy the following criteria:
\begin{enumerate}
    \item The term should be semantically associated with the concept from the original query;
    \item The term extracted from Freebase should also be widely adopted in the Twitter corpus while talking about the concept;
    \item As the user's intent may change and events related to the given topic will develop over the time,
        the ranking function should favor the short-term words that are mostly used in recent tweets.
\end{enumerate}
The candidate terms extracted from Freebase meet the first criterion to some extent.
In order to satisfy the second criterion,
we score the candidate terms with an association based method on the basis of the 
top ranked $N$ pseudo-relevance documents (PRD):
\begin{equation}
    Score(w) = \sum_{D\in PRD} P(D)\cdot P(w|D)\cdot \prod_{i=1}^{n}P(q_i|D)
    \label{rm}
\end{equation}
where $P(D)$ is the document prior which is usually assumed to be uniform,
and $\prod_{i=1}^{n}P(q_i|D)$ is the query likelihood given the document model,
which is traditionally computed using Dirichlet smoothing.
To meet the third criterion,
we follow the work of \cite{li2003time}
and incorporate the temporal evidence into the document prior in Eq.\ref{rm}
by using an exponential distribution:
\begin{equation}
    P(D|T_D) = r\cdot e^{-r(T_Q-T_D)}
    \label{time}
\end{equation}
where $r$ is the exponential parameter that controls the temporal influence,
$T_Q$ is the query issue time and $T_D$ is the tweet post time.
Both $T_Q$ and $T_D$ are measured in fractions of days.
Note that $T_D$ is constantly less than $T_Q$ as we cannot use the future evidence.

Finally, we select the top scored $K$ words from the common description and domain specific properties, 
to form the knowledge query $Q_{fb}$ along with the terms extracted from meta properties.  
These $K$ words along with the ones from meta properties are treated equally and combined to form the knowledge query $Q_{fb}$.

\subsection{Mixture Feedback Model}
With the knowledge query environment,
we believe the information need is more understandable, which could lead to a
high precision in top retrieved tweets.
Based on this hypothesis,
we further utilize a model-based feedback to update the query representation.
More specifically,
we update the $\hat\theta_{Q_1}$ with the simple mixture model $\hat\theta_{F}$
which is widely used in microblog retrieval \cite{zhai2011mbfb,liang:jcdl12}.
\begin{equation}
    P(w|\hat\theta_{Q_2}) = (1-\beta) \cdot P(w|\hat\theta_{Q_1}) + \beta \cdot P(w|\hat\theta_{Q_{F}})
    \label{interpolate2}
\end{equation}
where $\beta \in [0,1]$ is a weighting parameter to control the amount of model-based feedback.

The model-based feedback model generates a feedback document by mixing the query topic model $\hat\theta_{F}$
with the collection language model $\hat \theta_C$.
Under this simple mixture model,
the log-likelihood of feedback documents $F$ is:
\begin{multline}
\log{P(F|\hat\theta_{F})} = \sum_{w } {c(w,F})  \cdot  \\ 
		\log(  (1-\lambda) \cdot P(w|\hat\theta_{F})+\lambda \cdot P(w|\hat \theta_C) )
\end{multline}
where $c(w,F)$ is the count of word $w$ occurred in the set of feedback documents $F$.
Then we follow the work of \cite{zhai2011mbfb} and implement the EM algorithm
with the fixed smoothing parameter $\lambda=0.5$.
No matter whether or not the query finds its knowledge terms in Freebase,
the query environment will be updated by the model-based feedback.

\section{Evaluation}
\subsection{Experimental Setup}
In this section, we describe the experimental dataset and evaluation methods
which are adopted in TREC Microblog Track \cite{ounis:trec11,ian:trec12,jimmy:trec13}.
In addition, baselines are set up to estimate the effect of the proposed methods.
Notations and abbreviations that appear in our experiments are given in Table \ref{tab:notation}.
\begin{table}[htbp!]
    \centering
    \caption{Abbreviations of Experimental Systems.}
    \begin{tabular}{|p{66pt}|p{160pt}|}
        \hline
        \textbf{Abbreviation} & \textbf{Description} \\
        \hline
        \textbf{SimpleKL} & Simple KL-divergence retrieval model without query expansion and document expansion. \\
        \hline
        \textbf{QESMM} & KL-divergence retrieval model with model-based feedback \cite{zhai2011mbfb}. \\
        \hline
        \textbf{QEWiki} & KL-divergence retrieval model with query model $\hat\theta_{Q_1}$, and expansion terms are derived from top retrieved Wikipedia articles. \\
	  \hline
         \textbf{RTRM} & Real-time ranking model proposed in \cite{liang:jcdl12}, using a two-stage query expansion method  and gaussian function based temporal re-ranking with ranking position profile.\\
         \hline
        \textbf{QEFB} & KL-divergence retrieval model with query model $\hat\theta_{Q_1}$, and terms are derived from both description property and meta properties in Freebase.\\
        \hline
        \textbf{QEFBNT} & KL-divergence retrieval model with query model $\hat\theta_{Q_1}$ without
        temporal prior while selecting knowledge terms.\\
        \hline
        \textbf{QEManualFB} & The same as \textbf{QEFB} except that the Freebase concepts are manually selected.\\
        \hline
        \textbf{QEFB+SMM} & KL-divergence retrieval model with query model $\hat\theta_{Q_2}$.\\
        \hline
    \end{tabular}
    \label{tab:notation}
\end{table}

\subsubsection{Data Set}
Two corpora (i.e. Tweets11 and Tweets13 collection) are used in our experiments.
Instead of distributing the microblog corpus via physical or direct downloading, 
TREC organizers release a streaming API \footnote{https://github.com/lintool/twitter-tools} to participants \cite{jimmy:trec13}. 
Using the official API, we crawled a set of local copies of the canonical corpora.
Tweets11 collection has a sample of about 16 million tweets, ranging from January 24, 2011 to February 8, 2011 
while Tweets13 collection contains about 259 million tweets, ranging from February 1, 2013 to March 31, 2013. 
In addition, we also crawled all the shortened URLs contained in Tweets11 and Tweets13 Corpora, 
and inferred their topic information (i.e. title of the crawled webpage) to enrich the original tweets. 
In particular, we consider the title information of the embedded URLs as the local context of the original tweets 
and combine it with the original tweets to form the tweet language model  \cite{liang:jcdl12}. 
Tweets11 is used to evaluate the effectiveness of the proposed real-time Twitter search systems 
over 50 official topics (MB001-MB050) in the TREC'11 Microblog track as well as 60 official topics (MB051-MB110) in the TREC'12 Microblog track, 
respectively\footnote{The topic numbered MB050 and MB076 has no relevant tweets. Therefore, we did not use them for our experiments.}.
And, Tweets13 is used in evaluating the proposed real-time Twitter search systems 
over 60 official topics (MB111-MB170) in the TREC'13 Microblog track.
In our experiments, TREC'11 topics  are used for tuning the parameters 
and then we use the best parameter settings to evaluate our methods with TREC'12  and TREC'13 topics.

The tweets and their corresponding topic information were preprocessed in several ways.
We first discarded the non-English tweets using a language detector with infinity-gram,
named \textit{ldig} \footnote{http://github.com/shuyo/ldig}.
Second, in conformance with the track's guidelines, 
all simple retweets were removed by deleting documents beginning with the string `RT'.
Moreover, each tweet was stemmed using the Porter algorithm
and stopwords were removed using the InQuery stopwords list.
\subsubsection{Evaluation Metric}
In TREC Microblog Track, 
tweets were judged on the basis of the defined information using a three-point scale \cite{ounis:trec11}:
irrelevant (labeled as 0), minimally relevant (labeled as 1), and highly relevant (labeled as 2). 
The main evaluation metric is Mean Average Precision (MAP) for top $1000$ documents
and Precision at N (P@N), which are widely used in IR.
MAP and P@30 with respect to \textit{allrel} (i.e. tweet set judged as highly or minimally relevant) are used in this paper.
We also do a query-by-query analysis and conduct t-test to determine 
whether the improvements on MAP and P@30 are statistically significant.

\subsubsection{Baselines}
To demonstrate the performance of our proposed method,
we compare our knowledge-based query expansion methods with several baseline methods. 

(1) The simple KL-divergence retrieval model (denoted as \textbf{SimpleKL}) \cite{zhai2001study} is used as our first baseline.
That is, we estimate $\hat\theta_Q$ and $\hat\theta_D$ with empirical word distribution,
and we choose Dirichlet smoothing method for document model estimation.
Throughout this paper, we set the Dirichlet smoothing parameter $\mu = 100$,
which has been reported for a good retrieval performance in microblog retrieval \cite{liang:jcdl12}. 

(2) We use the Simple Mixture Model \cite{zhai2011mbfb} (denoted as \textbf{QESMM}) as our second baseline,
and optimize the number of feedback documents to $7$ and the number of terms in the feedback model to $5$.
The smoothing parameter $\beta$ is set as $0.9$. 

(3) \textbf{QEWiki} is a Wikipedia-based query expansion method, which is similar with the work of \cite{li2007improving}.
We downloaded a local copy of Wikipedia data for faster access and indexed the articles using Lemur toolkit \footnote{http://www.lemurproject.org/lemur.php} (version 4.12).
The expansion terms are derived from top ranked Wikipedia articles.
In our experiments, we rank Wikipedia articles using language model (i.e. SimpleKL), 
and total $10$ terms are picked from the top $5$ documents.
Then we treat the terms as a new query and interpolate it with the original query.
The interpolation parameter $\alpha$ in Eq.\ref{interpolate} for \textbf{QEWiki} is set as $0.4$. 

(4) We also compare our method with the state-of-the-art real-time ranking model (denoted as \textbf{RTRM}) 
under language modeling framework, proposed by \cite{liang:jcdl12}.
\textbf{RTRM} approach also utilized a two-stage pseudo-relevance feedback query expansion to estimate the query language model.
Besides, \textbf{RTRM} adopts a temporal re-ranking component to evaluate the temporal aspects of tweets.

We tune all the parameters of these models with TREC'11 topics on Tweets11 corpus.

\subsection{Experimental Results}
We conduct several experiments to measure the effects of our query expansion methods.
For our knowledge-based query expansion method,
we label the method with query model $\hat\theta_{Q_1}$ as \textbf{QEFB},
and the one with $\hat\theta_{Q_2}$ as \textbf{QEFB+SMM}.
When selecting knowledge terms from Freebase description and domain specific (e.g. Business domain) properties,
we set the top ranked PRD number $N$ to $100$
and the expanded term number $K$ to $5$.
The exponential parameter $r$ for temporal prior is set as $0.1$.
$\alpha$ in Eq.\ref{interpolate} is set as $0.5$,
which means we regard the original query and the knowledge query equally important.
The query expansion parameters in the mixture feedback model are set like \textbf{QESMM} 
except that the interpolation $\beta$ is set as $0.6$.
All the parameters are tuned with TREC'11 topics. 
Then we test the optimized models with TREC'12 and TREC'13 topics.

Table \ref{tab:qe} shows the performance comparison of different query expansion methods.
For statistical significance, we used a paired t-test. 
$\dag$, $\ddag$, $\P$ and $\S$  indicate that the corresponding improvements over 
\textbf{SimpleKL}, \textbf{QESMM}, \textbf{QEWiki} and \textbf{RTRM} are statistically significant ($p<0.05$), respectively.
Note that all the methods listed in the table estimate the document model as \textbf{SimpleKL}.
As we can see, all of the query expansion methods have significant MAP and P@30 improvements
compared with the \textbf{SimpleKL} method,
which indicates the effectiveness of query expansion in microblog retrieval.
Besides, \textbf{QEFB} performs better than the Wikipedia-based query expansion method \textbf{QEWiki}.
This shows the superiority of our Freebase-based query expansion method and demonstrate the effectiveness of the structured data.

When the query is expanded with the Freebase knowledge query,
our approach can retrieve more relevant documents in the top results.
Thus, we can further improve the retrieval performance by
combining the knowledge-based expansion method with mixture feedback model.
Our knowledge-based query expansion method \textbf{QEFB+SMM} achieves the best retrieval performance 
in the three topic sets with respect to both MAP and P@30 metrics.
More specifically, for TREC'12 topics, our method \textbf{QEFB+SMM}  improves the MAP over \textbf{SimpleKL} and \textbf{QESMM} by 23.80\% and 12.35\%, respectively; 
while the corresponding increments in terms of P@30 are 14.91\% and 11.56\%, respectively.
For TREC'13 topics, the \textbf{QEFB+SMM} raises the MAP over \textbf{SimpleKL} and \textbf{QESMM} 
by 24.81\% and 12.02\%,respectively; 
while the corresponding P@30 improvements are 16.42\% and 12.37\%, respectively.
Moreover, Our method also beats the state-of-the-art baseline \textbf{RTRM}, which uses a two-stage query expansion method.
\begin{table*}[htbp]
    \center
    \caption{The performance comparison of different query expansion methods. The best performances are marked in bold.}
    \begin{tabular}{|l|l|l|l|l|l|l|}
	 \hline
        \textbf{Topics} & \multicolumn{2}{c|}{\textbf{TREC'11}} & \multicolumn{2}{c|}{\textbf{TREC'12}} & \multicolumn{2}{c|}{\textbf{TREC'13}} \\
        \hline
        \textbf{Method} &  \textbf{MAP} & \textbf{P@30} &  \textbf{MAP} & \textbf{P@30}  &  \textbf{MAP} & \textbf{P@30}  \\
        \hline
	\textbf{SimpleKL}		& 0.3645 &	0.3850  &	0.2727	&	0.3938	 &	0.2926	&	0.4939\\
	\textbf{QESMM}		& 0.3957 &	0.4218 &	0.3005	&	0.4056	 &	0.3260	&	0.5117\\
	\textbf{QEWiki}			& 0.4041 & 0.4177  &    0.3175   &    0.4203   &   0.3099   &   0.5111 \\
	\textbf{RTRM}	     	      &0.4226 & 0.4463  &	0.3250	&	0.4458	 &	0.3507	&	0.5406\\
	\hline
	\textbf{QEFB}			&0.4289$\dag$ & 0.4252  &	0.3198$\dag$	&	0.4333	&		0.3149$\dag$ &	0.5117\\
	\textbf{QEFB+SMM}	& \textbf{0.4369}$\dag \ddag \P$ & \textbf{0.4497}$\dag \P$  &	\textbf{0.3376}$\dag \ddag$	&	\textbf{0.4525}$\dag \ddag \P$	 &	\textbf{0.3652}$\dag \ddag \P$	&	\textbf{0.5750}$\dag \ddag \P \S$\\
        \hline
    \end{tabular}
    \label{tab:qe}
\end{table*}

To further demonstrate the effectiveness of our proposed method, 
we also compare our \textbf{QEFB+SMM} with the top three automatic runs in TREC 2012 and 2013 Microblog track.
Table \ref{tab:trecbest} shows the MAP and P@30 performances of all these runs.
Note that for TREC'12, the ranking scores are computed with respect to the \textit{highrel} set \cite{ian:trec12};
while for TREC'13, the scores are computed in the \textit{allrel} set \cite{jimmy:trec13}.
From the table, we can observe that our system is comparable with the top three runs in TREC Microblog track.
Moreover, \textbf{QEFB+SMM} even beats the best automatic run in TREC'13 with respect to both evaluation metrics.
\begin{table}[htbp]
    \center
    \caption{The performance comparison of our \textbf{QEFB+SMM} with TREC best runs. The best performances are marked in bold.}
    \begin{tabular}{|l|l|l|l|l|l|l|}
	 \hline
        \textbf{Topics} &   \multicolumn{2}{c|}{\textbf{TREC'12}} & \multicolumn{2}{c|}{\textbf{TREC'13}} \\
        \hline
        \textbf{Method} &  \textbf{MAP} & \textbf{P@30}  &  \textbf{MAP} & \textbf{P@30}  \\
        \hline
	\textbf{1st run}		&	\textbf{0.2642}	&	\textbf{0.2701} &	0.3524	&	0.5528\\
	\textbf{2nd run}	&	0.2411	&	0.2446 &	0.3506	&	0.5544\\
	\textbf{3rd run}	     	&	0.2093	&	0.2384	&	0.3494	&	0.5372 \\
	\hline
	\textbf{QEFB+SMM}		 & 0.2415	&	0.2429  &	\textbf{0.3652}	&	\textbf{0.5750}\\
        \hline
    \end{tabular}
    \label{tab:trecbest}
\end{table}

\subsection{Discussion}
\label{sec:Discussion}
Many parameters in our proposed approach can affect the system performance.
In this section, we analyze the robustness of the parameter settings in knowledge-based query expansion method.
All these experiments in this section are run on TREC'11 topics, 
which are used for parameter selection.

\subsubsection{Effects of Knowledge Query}
For the query modeling,
we propose using knowledge query to make the information need more comprehensible.
Many factors affect the quality of the knowledge terms:
(1) whether the maximum match algorithm can get topic-related concept from the Freebase;
(2) the number of knowledge terms $K$ and
(3) the number of pseudo-relevance documents $N$ used for term selection.

\begin{figure}[htbp]
    \centering
    \includegraphics[width=0.232\textwidth]{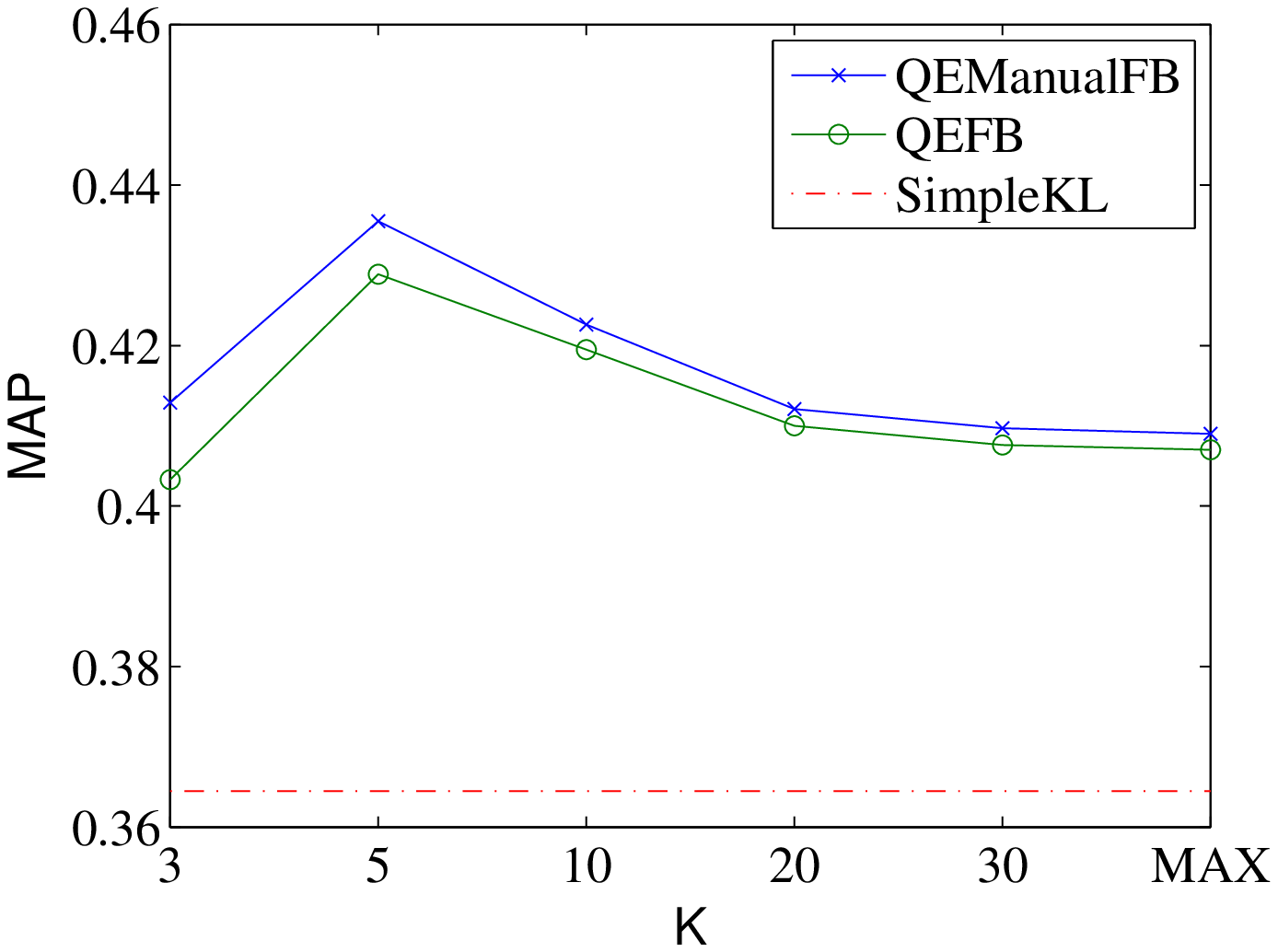}
    \includegraphics[width=0.232\textwidth]{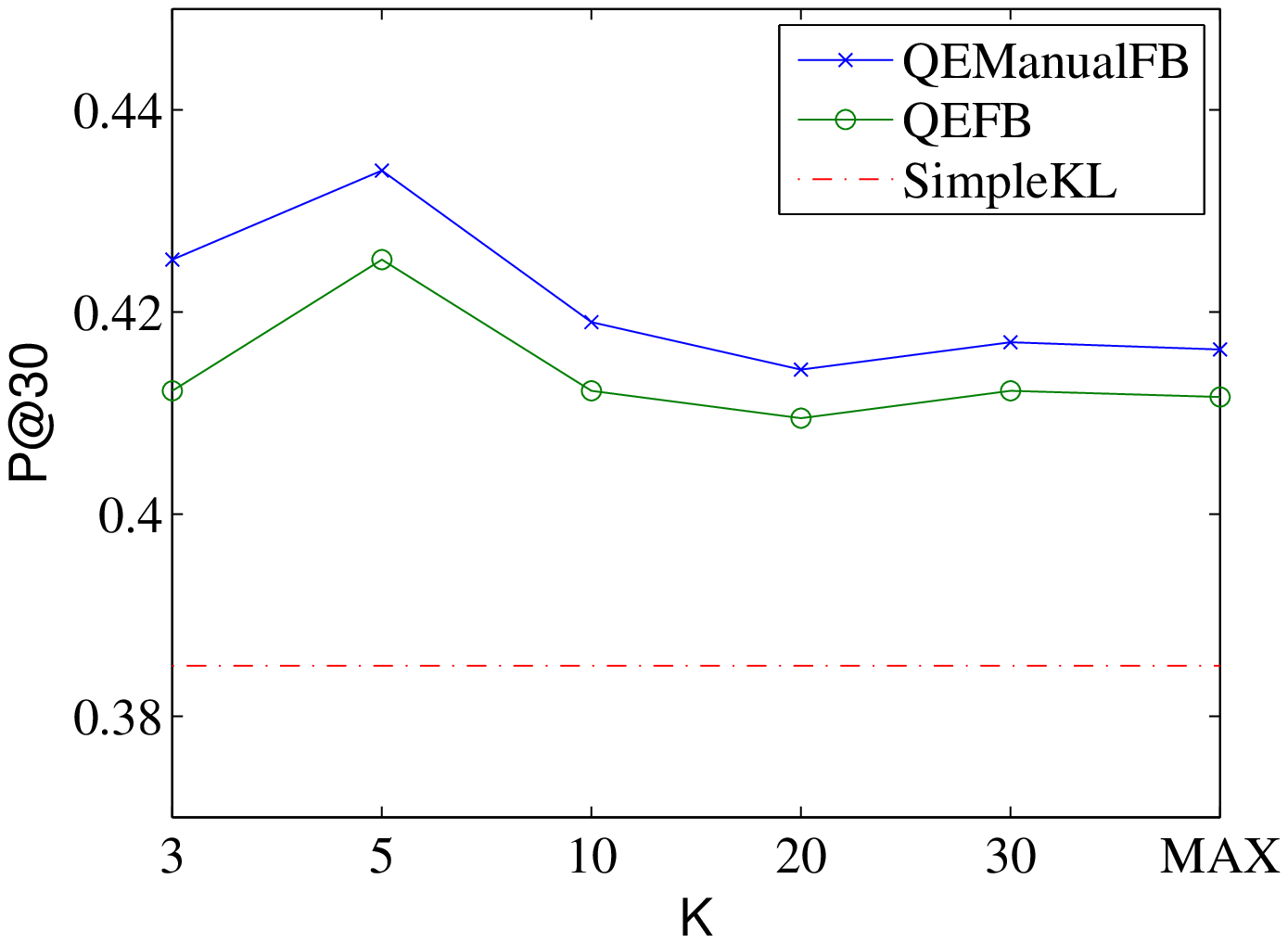}
    \caption{Sensitivity to the selected knowledge term number $K$.}
    \label{fig:termcount}
\end{figure}

To answer the first question, we create the run \textbf{QEManualFB},
which means we manually select the concept from Freebase for each query.
The interpolation parameter $\alpha$ for all these models are set as $0.5$.
Figure \ref{fig:termcount} shows the MAP and P@30 scores of all the models
for different $K$ and fixed $N=100$.
In particular, $MAX$ means all the candidate terms that satisfy $Score(w) > 0$ in Eq.\ref{rm} are selected.
We can see that though \textbf{QEFB} is not better than \textbf{QEManualFB},
the performance gap between them is not large,
which verifies the effectiveness of our concept match algorithm.
Moreover, when $K$ is set around $5$,
\textbf{QEFB} can get its optimal retrieval performance and is significantly better than that of \textbf{SimpleKL},
which indicates the effectiveness of the association based term selection method.

\begin{figure}[htbp]
    \centering
    \includegraphics[width=0.232\textwidth]{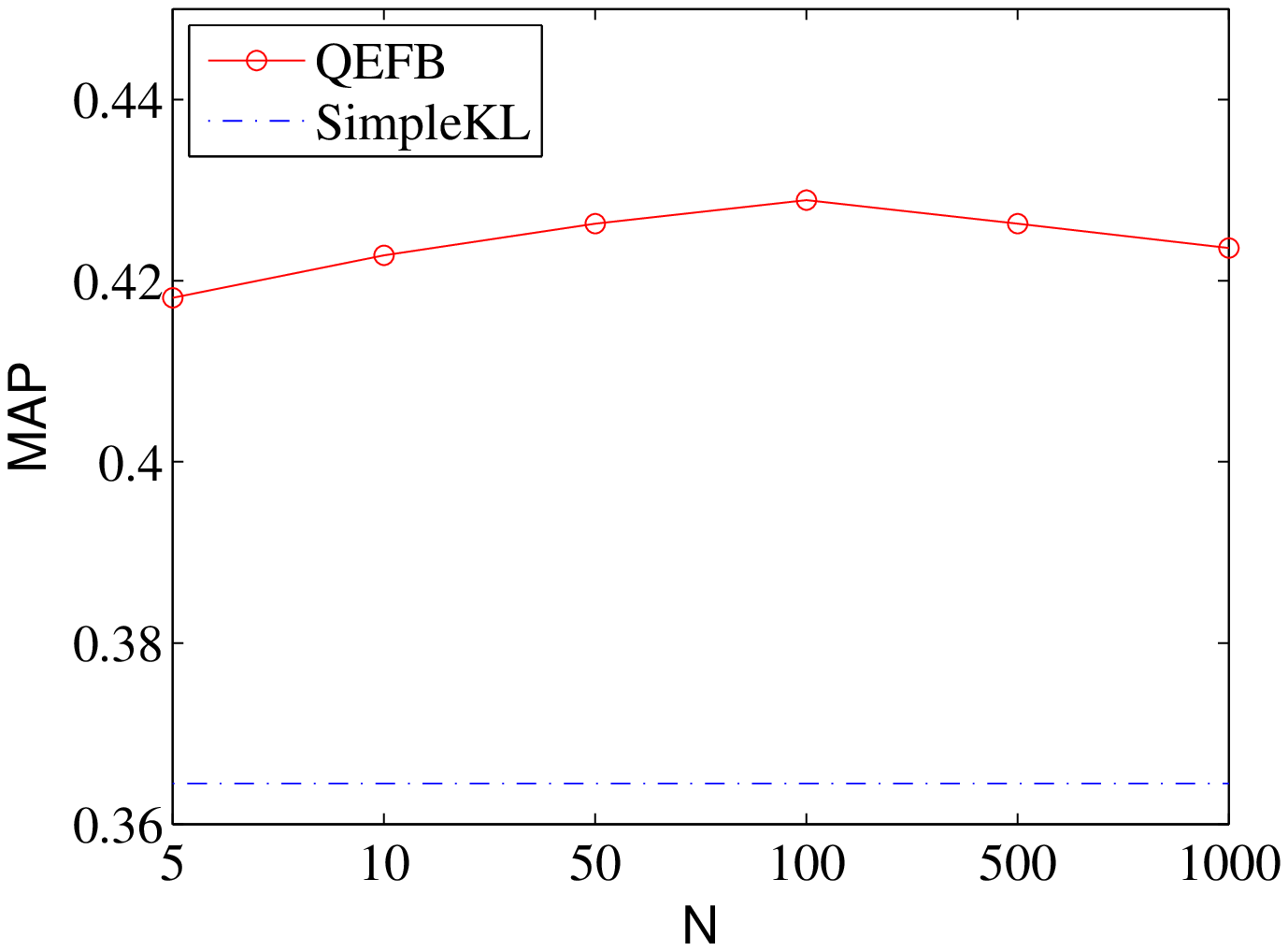}
    \includegraphics[width=0.232\textwidth]{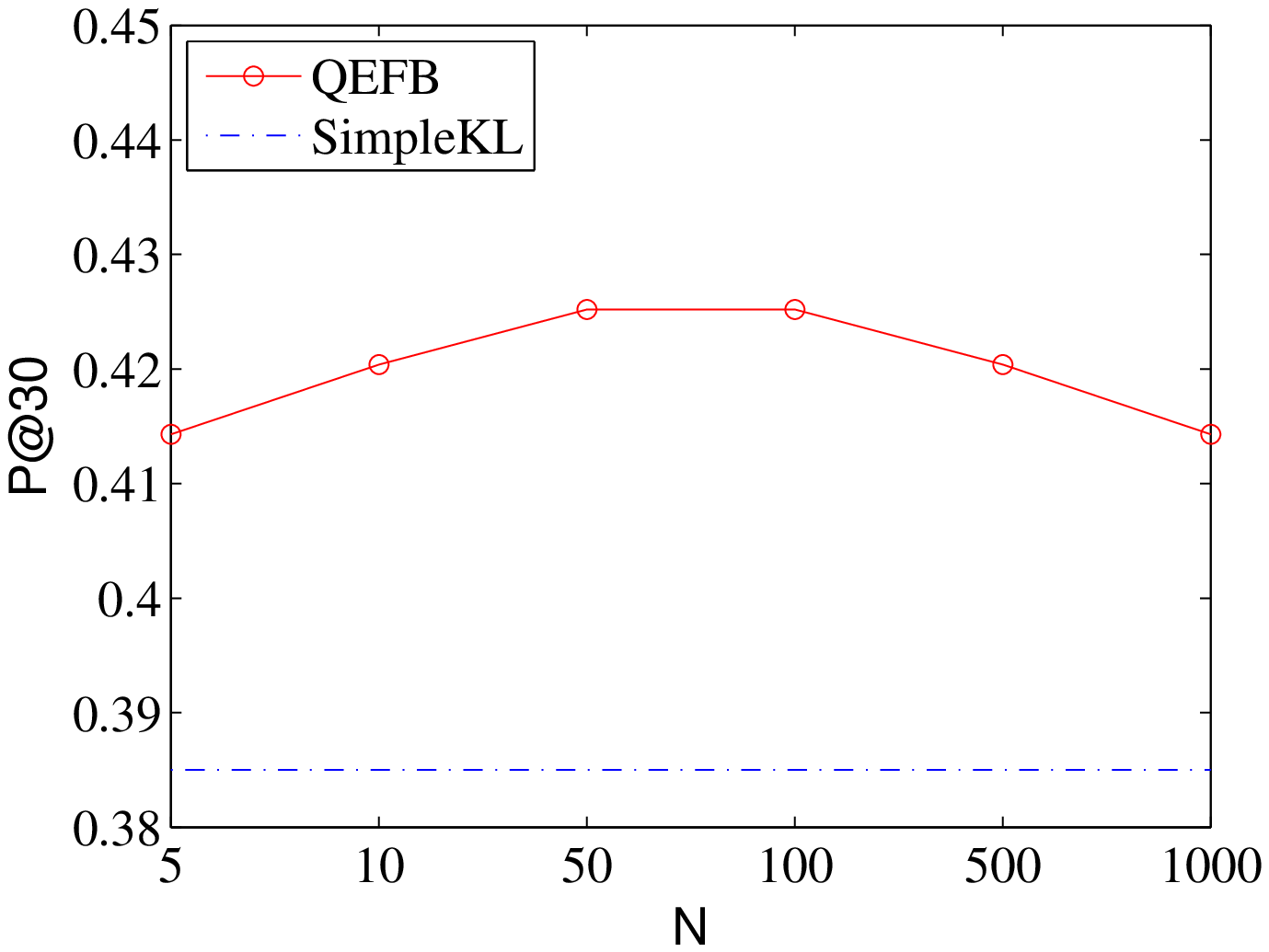}
    \caption{Sensitivity to the PRD number $N$ for knowledge term selection.}
    \label{fig:doccount}
\end{figure}

To further show the parameter sensitivity to the PRD number for term selection,
we fix the term number $K$ as $5$ and change the PRD number $N$.
Figure \ref{fig:doccount} shows the MAP and P@30 scores of our \textbf{QEFB} model against different values of $N$.
It is readily apparent that \textbf{QEFB} can achieve its optimal performance when $N$ is set to $100$.
That is, top $100$ pseudo-relevance documents can provide adequate information
for selecting good knowledge terms from Freebase description and domain specific properties.

\subsubsection{Effects of Temporal Evidence}
In the previous work \cite{efron:sigir11},
it was shown that the selection of the rate parameter $r$ for the exponential distribution
when applying temporal prior has a strong effect on retrieval.
In the previous sections, we set $r$ in Eq.\ref{time} as $0.1$. 
Now, we want to verify the effect of the temporal evidence in our expansion methods.

\begin{figure}[htbp]
    \centering
    \includegraphics[width=0.49\textwidth]{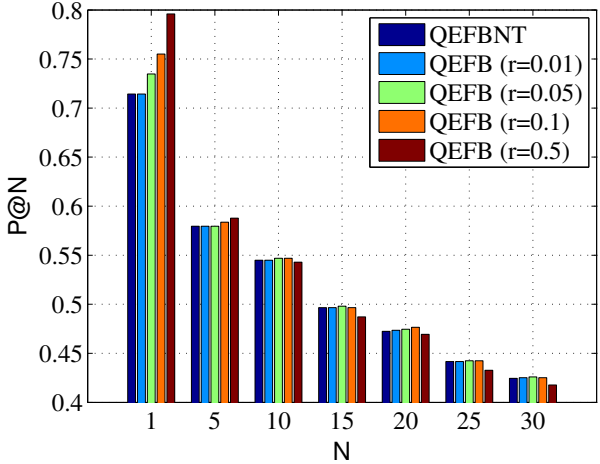}
    \caption{Sensitivity of the \textbf{QEFB} model to the exponential rate parameter $r$.}
    \label{fig:qetime}
\end{figure}

In our method,
the temporal prior affects the knowledge terms selected from the Freebase properties.
A large $r$ favors the terms that are used recently in the pseudo-relevance documents.
For better comparison, we create a run named \textbf{QEFBNT} ignoring the temporal evidence.
Figure \ref{fig:qetime} shows the P@N scores of \textbf{QEFB} with different values of $r$.
Only four values of $r$ are shown here, although more were tried.

We can observe from the figure that an appropriate $r$ can improve the retrieval performance 
compared with the \textbf{QEFBNT} in terms of P@N.
Besides, a large $r$ can highly improve the precision of top retrieved tweets.
Note that \textbf{QEFB}  ($r=0.5$) has maximum P@1 and P@5 scores compared with other settings.
However, it does not show any superiority over other models with respect to the P@N $(N \geq 10)$ scores.
In fact, the MAP score of \textbf{QEFB} ($r=0.5$) is also lower than \textbf{QEFB} with a small $r$.
A rational explanation for this interesting phenomenon may be that, with more short-term words,
more tweets with higher relevance can be retrieved easily and thus the precision of top ranked $5$ tweets is boosted.
But at the same time, more irrelevant tweets in top $30$ documents could be retrieved as these terms overemphasize the recency.

Taking the query ``water shortage'' (MB111) as an example, 
the top knowledge words of \textbf{QEFBNT} are ``affect, global, area''.
For the \textbf{QEFB} of $r=0.5$, the top words are  ``africa, drought, play''.
This indicates that people mainly focus on drought in
Africa recently when they are talking about the water shortage.
In our system, we finally choose \textbf{QEFB} ($r=0.1$) which has both high and stable P@N ($1 \leq N \leq 30$) and MAP scores.

\subsubsection{Effects of the Interpolation Coefficients}
\begin{figure}[htbp]
    \centering
    \includegraphics[width=0.232\textwidth]{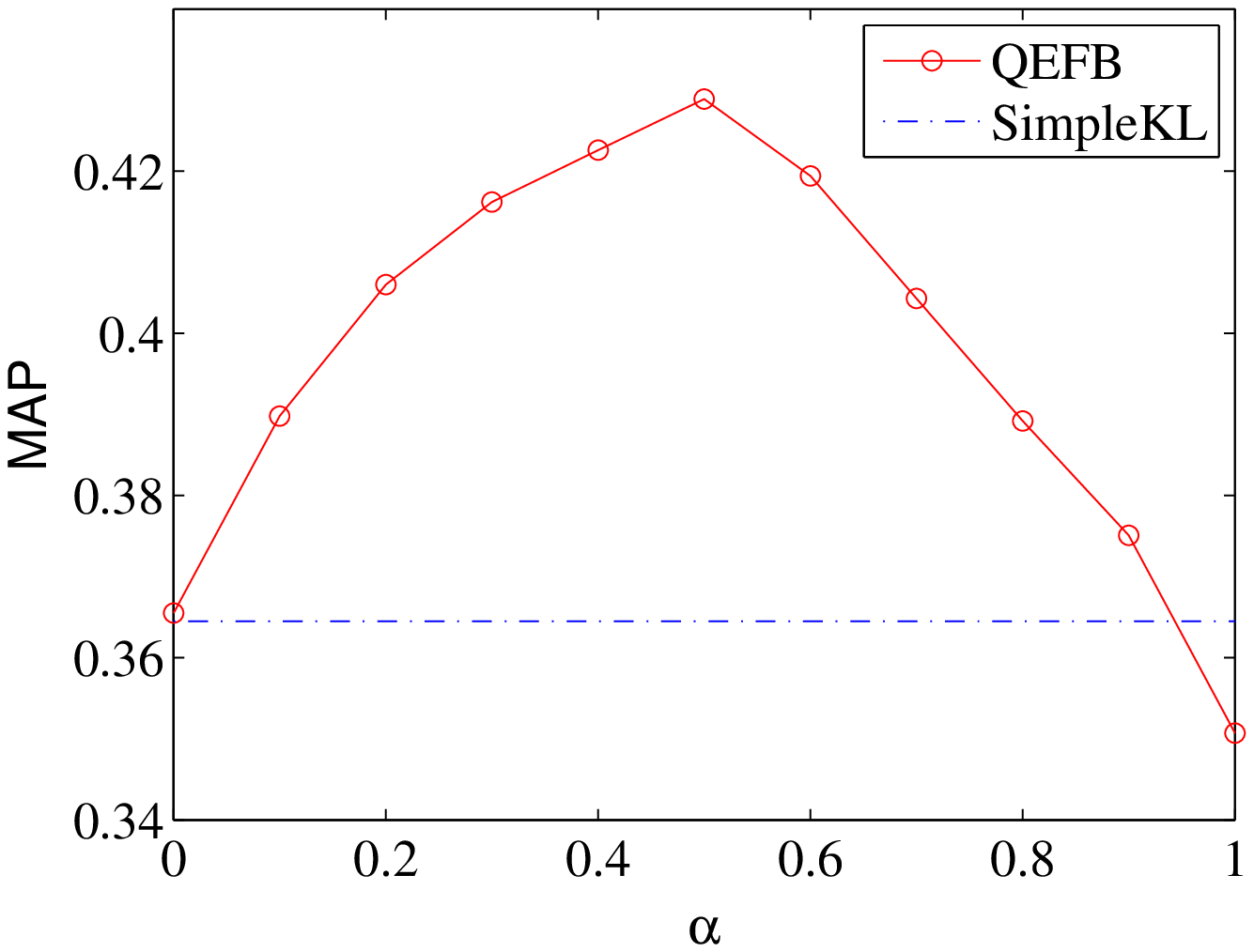}
    \includegraphics[width=0.232\textwidth]{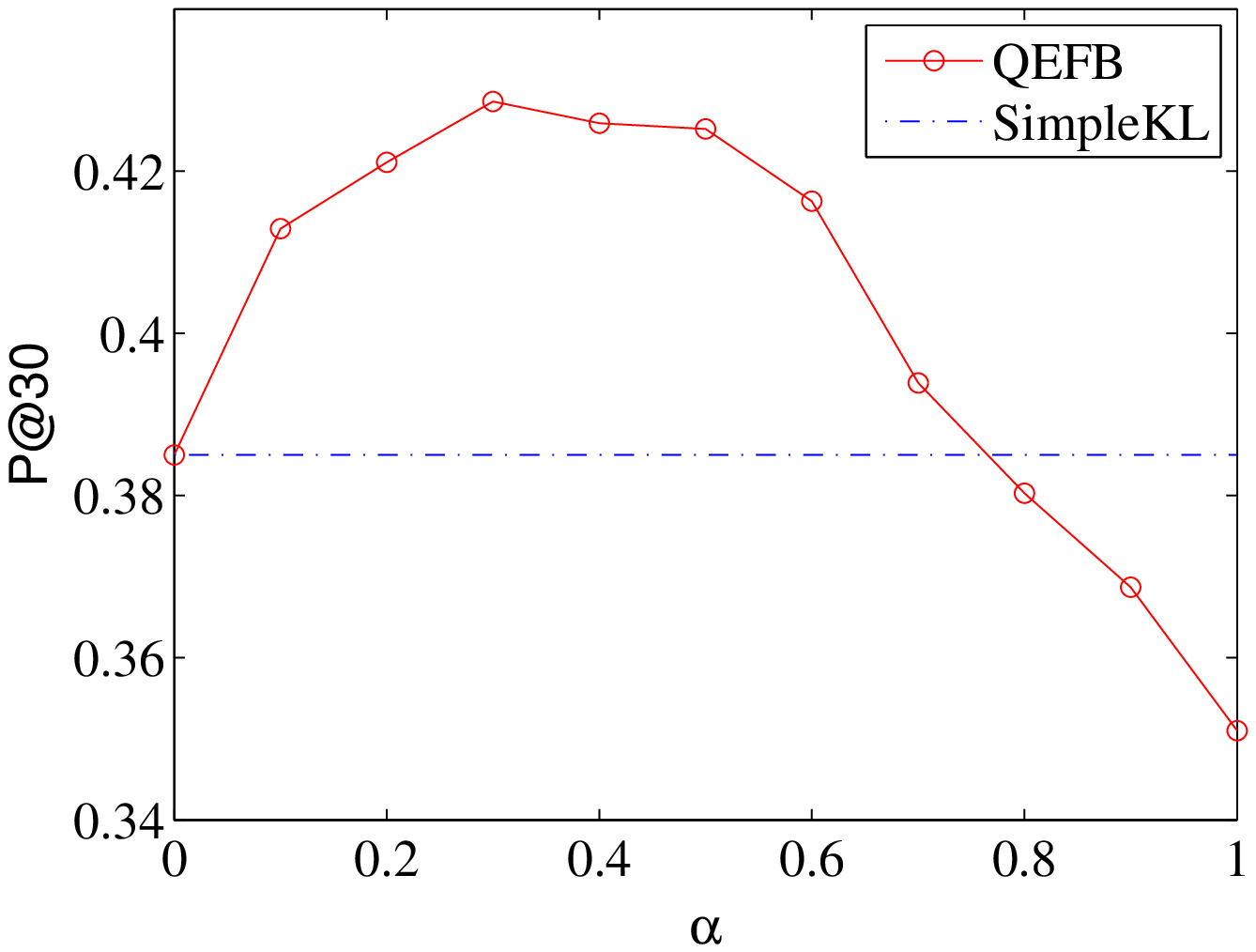}
    \caption{Sensitivity to the first-stage knowledge query expansion coefficient $\alpha$.}
   \label{fig:qealpha}
\end{figure}

Recall that we first expand the query with knowledge query, 
and further expand the updated query $\hat\theta_{Q_1}$ with model-based feedback.
The first-stage query expansion is controlled by a coefficient $\alpha$, 
while the second-stage expansion is controlled by $\beta$.
Figure \ref{fig:qealpha} shows the performance variance of \textbf{QEFB} ($N=100, K=5$) 
against different values of $\alpha$.
When $\alpha=0$, \textbf{QEFB} degenerates into the baseline method \textbf{SimpleKL}.
When $\alpha=1$, we completely ignore the original query and only use the knowledge query.
We can observe that the performance of \textbf{QEFB} is better than \textbf{SimpleKL} 
when $\alpha$ is no greater than $0.7$.
The optimal performance can be obtained when $\alpha$ is set around $0.5$.

\begin{figure}[htbp]
    \centering
    \includegraphics[width=0.232\textwidth]{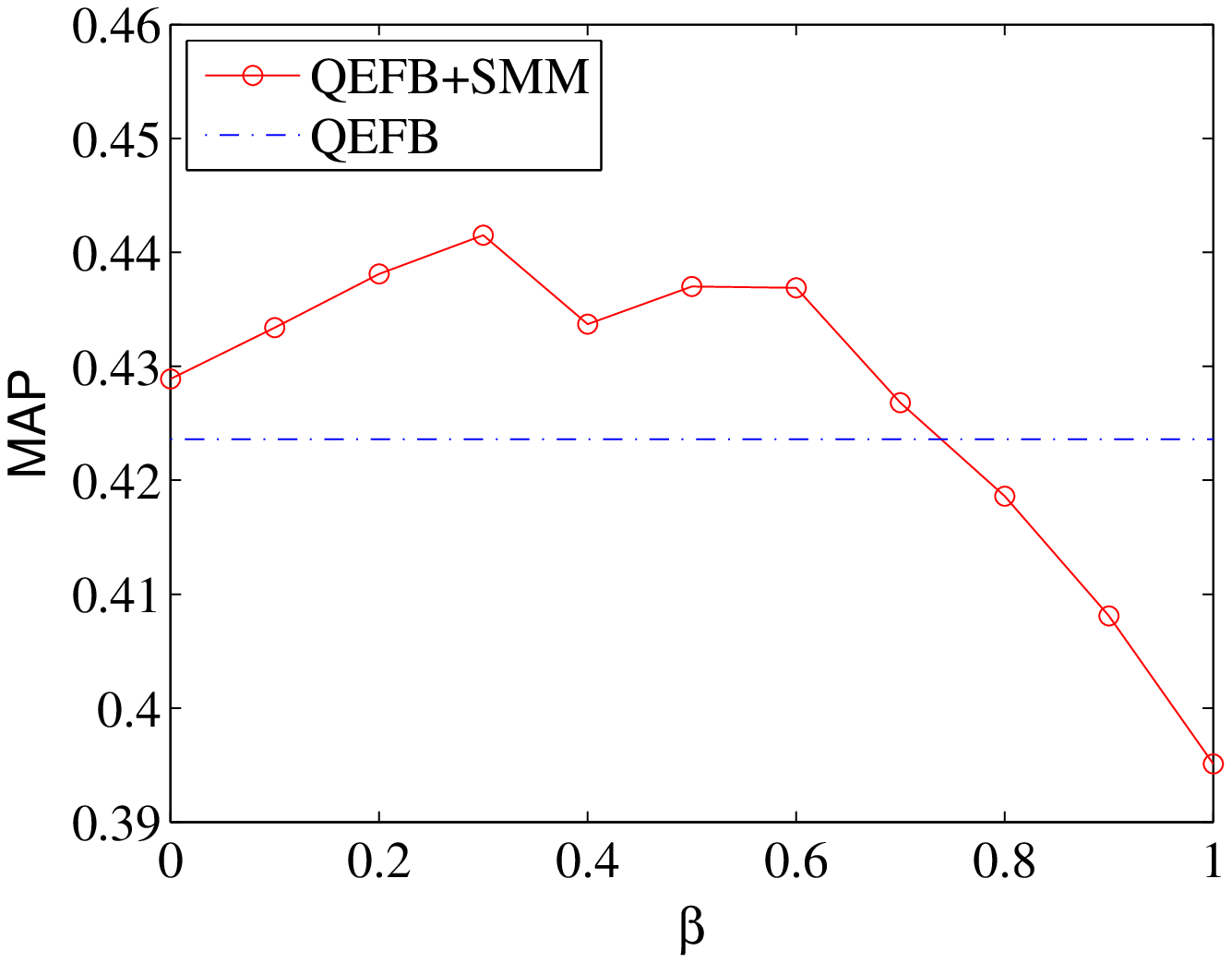}
    \includegraphics[width=0.232\textwidth]{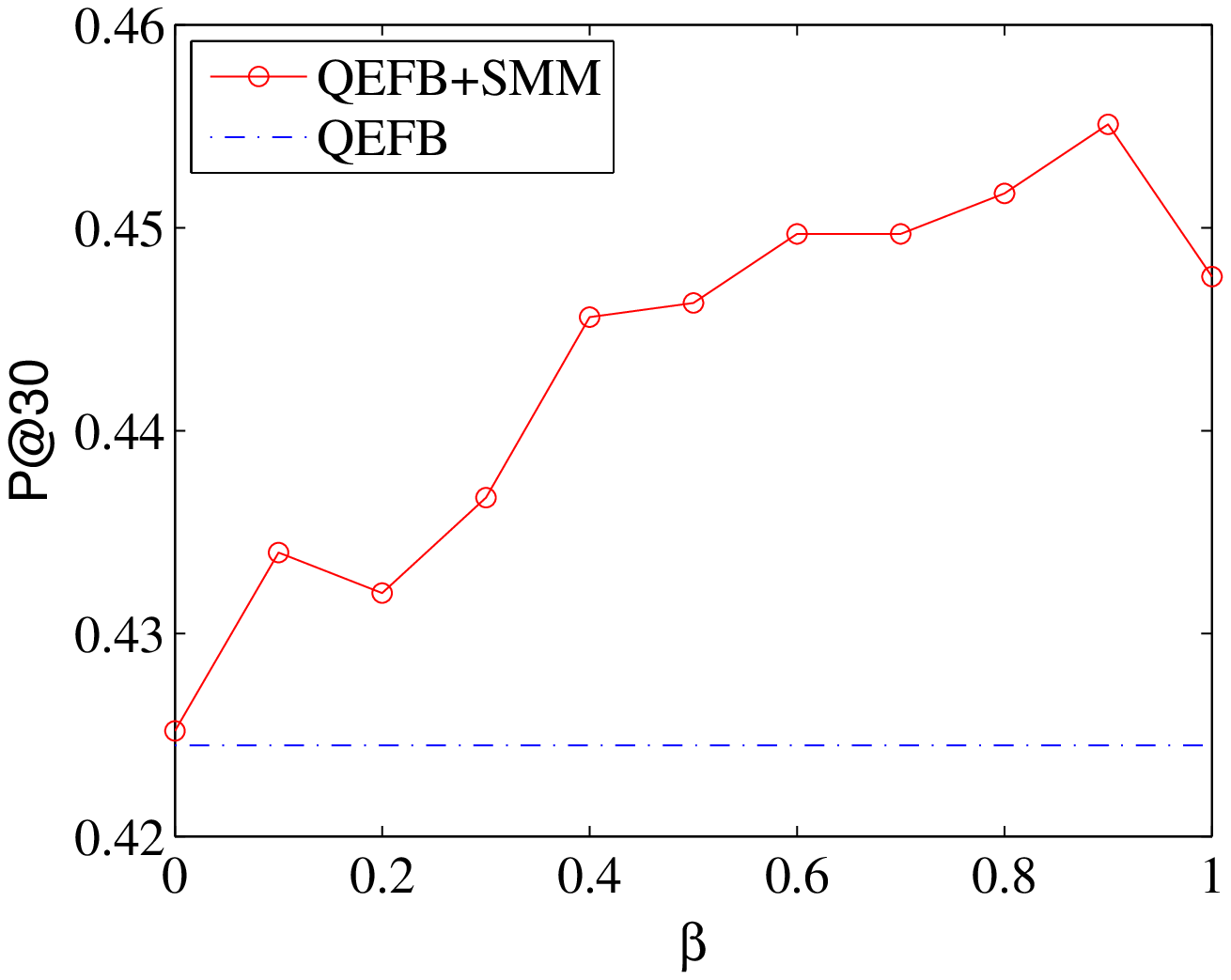}
    \caption{Sensitivity to the second-stage mixture feedback interpolation coefficient $\beta$.}
   \label{fig:qebeta}
\end{figure}

Figure \ref{fig:qebeta} shows the performance variance of \textbf{QEFB+SMM} against different values of $\beta$.
When $\beta=0$, the \textbf{QEFB+SMM} degenerates into \textbf{QEFB}.
The second-stage expansion seems to be more robust and constantly better than \textbf{QEFB} with respect to P@30.
After knowledge-based query expansion, the query can be more comprehensible and get more top related tweets,
which leads to further improvement with traditional model-based feedback. 
However, when it comes to the MAP metric, 
the performance of \textbf{QEFB+SMM} drops when $\beta$ is larger than $0.3$.
Finally, we choose $\beta = 0.6$ which is a tradeoff between MAP and P@30.

\section{Conclusion and Future Work}
In this study, we proposed using knowledge-based query expansion 
to solve the problems in microblog search.
With the knowledge terms derived from the Freebase, the queries in microblogosphere
can be more comprehensible and thus more relevant documents can be retrieved.
The knowledge terms from Freebase should co-occur with query terms in PRD, 
which has the potential to alleviate the topic drift induced by knowledge-based QE.
Freebase's structured information is well utilized in knowledge query generation procedure.
Moreover, we incorporated the temporal evidence into query representation.
Thus the proposed method favors recent tweets
which satisfies the real-time information need in microblog retrieval.
Our thorough evaluation, using two standard TREC collections, 
demonstrates the effectiveness of the proposed method.

Many studies remain for the future work.
One of the most interesting directions is to explore more complicated algorithms to explore the domain information of Freebase.
By further analyzing the domain information of the concepts for a given query,
we can also assign the retrieved tweets to different domains, 
which can be used to generate a structural result representation.
Moreover, we can classify the queries into two categories as
temporal-dependent and temporal-independent ones, 
and use different strategies to estimate temporal evidence for each category.

\section{Acknowledgments}
The work reported in this paper was supported by the National Natural science Foundation of China Grant 61370116.

\bibliographystyle{aaai}
\bibliography{aaai}  

\end{document}